\documentclass[11pt,a4paper]{article}

\usepackage{amsmath,amssymb,amsthm,amstext} 
\usepackage{a4,amsmath,amsfonts,amssymb} 
\usepackage[pdftex]{graphics} 
\usepackage{graphicx}
\usepackage{caption}

\renewcommand{\epsilon}{\varepsilon}
\renewcommand{\phi}{\varphi}
\theoremstyle{definition} \newtheorem{defi}{Definition}[section]
\theoremstyle{plain} \newtheorem{thm}[defi]{Theorem}
\theoremstyle{definition} \newtheorem{rem}[defi]{Remark}
\theoremstyle{definition} \newtheorem{exmpl}[defi]{Example}
\theoremstyle{definition} \newtheorem{algo}[defi]{Algorithm}
\theoremstyle{plain} \newtheorem{lem}[defi]{Lemma}
\theoremstyle{plain} \newtheorem{prop}[defi]{Proposition}
\theoremstyle{plain} \newtheorem{cor}[defi]{Corollary}
\theoremstyle{definition} \newtheorem*{mainres}{Main Results}
\theoremstyle{definition} \newtheorem*{danke}{Acknowledgement}

\newcommand{\R}{\mathbb{R}}
\newcommand{\Z}{\mathbb{Z}}

\newcommand{\n}{{(n)}}

\newcommand{\I}{\mathcal{I}}
\newcommand{\J}{\mathcal{J}}
\newcommand{\Scal}{\mathcal{S}}
\newcommand{\F}{\mathcal{F}}

\newcommand{\one}{{\mathbf{1}}}
\newcommand{\zero}{{\mathbf{o}}}

\DeclareMathOperator{\nest}{nest}
\DeclareMathOperator{\support}{support}
\DeclareMathOperator{\dasinnere}{interior}
\DeclareMathOperator{\codim}{codim}
\DeclareMathOperator{\tdeg}{tdeg}
\DeclareMathOperator{\tcoef}{tcoef}
\DeclareMathOperator{\tchar}{tchar}
\DeclareMathOperator{\td}{td}
\DeclareMathOperator{\tc}{tc}
\DeclareMathOperator{\clayton}{cl}
\DeclareMathOperator{\gumbel}{gu}
\DeclareMathOperator{\unifc}{cu}
\DeclareMathOperator{\tnest}{tnest}
\DeclareMathOperator{\NC}{NC}

\newcommand{\interior}[1]{\dasinnere({#1})}
\newcommand{\unitbox}[1]{[0,1]^{#1}}
\newcommand{\dset}[2]{\big\{ #1\,| \ \text{#2}\big\}}
\newcommand{\Dset}[2]{\Big\{ #1\,\big| \ \text{#2}\Big\}}

\newcommand{\tset}[2]{\{ #1\,| \ \text{#2}\}}

\begin{document}
\title{Shaping tail dependencies by nesting box copulas}
\author{Christoph Hummel \\Secquaero Advisors\footnote{Weinbergstr.~10, 
CH-8807 Freienbach, Switzerland, \texttt{christoph.hummel@secquaero.com}}}
\date{\today}

\hyphenation{co-pu-la di-men-sion-al}

\maketitle

\begin{abstract}
\noindent We introduce a family of copulas which are locally piecewise uniform 
in the interior of the unit cube of any given dimension.
Within that family, the simultaneous control of tail dependencies of all 
projections to faces of the cube is possible and we give an efficient sampling algorithm. 
The combination of these two properties may be appealing to 
risk modellers.
\end{abstract}
 
\section{Introduction} \label{sec:intro}

Copulas have become an accepted tool for modelling dependencies in the financial industry;
an overview from an applications point of view is given by 
P.~Embrechts in~\cite{Emb} together with a comprehensive 
list of references. One reason why risk modellers have become more used to working with copulas
is the fact that copulas allow the creation of models with increasing dependencies in the 
tails of the marginal distributions. 

We deal with two shortcomings of some of the currently used copula models. 
Firstly, practical algorithms 
for generating independent random samples of copulas efficiently, particularly in 
higher dimensions, are relatively scarce (refer to A.~McNeil et al.~\cite{QRM} 
and A.~McNeil~\cite{McN} for various simulation algorithms). 
Secondly, the parameters of the most prominent copulas such as the t-copulas, 
archimedean copulas (for instance Clayton and Gumbel) 
or nested archimedean copulas are related to pairwise dependencies of the 
marginal distributions so that a copula in the 
corresponding family is uniquely determined by the projections onto the 
$2$-dimensional faces. 

In this paper we give a general construction principle, which we call
\emph{tail nesting}, for copulas in any dimension~$r$. 
The characteristics of the tails can be shaped
by prescribing the tail dependencies in a very flexible manner.
The resulting copulas have efficient simulation algorithms.  

A copula $C$ corresponds to a Borel probability measure $c$ on the $r$-cube $\unitbox{r}$,
which, when projected to any $1$-dimensional face, yields the uniform probability measure. 
The correspondence between $C$ and $c$ is given via $C(u)=c([0,u])$ for $u=(u_1,\dots,u_n)$
and $[0,u]:=\prod_i[0,u_i]$. In the context of this paper, working directly with the measure 
$c$ turns out to be more convenient and we call $c$ as well a copula or a copula measure.
We refer to~\cite{QRM} or R.~Nelsen~\cite{Nel} for an introduction to copulas. 

Our main result is summarised below in this section. Some readers may prefer to read first the 
motivating examples in Section~\ref{sec:motivation} and then return to the paragraph below.

To begin with, we define the notion of tail dependency in higher dimensions which we 
work with. It is motivated by Example~\ref{exmpl:frac}, and the definition of 
lower tail dependency in~\cite{QRM}.

\begin{defi} \label{defi:tdep}
For $c$ as above, we define the \emph{tail degree} of $c$,
\begin{align}
\td(c)&\:=\inf\Dset{\tau}{$\liminf_{s\to 0}\frac{c(s\cdot\unitbox{r})}{s^\tau}=\infty$}\,. \\
\intertext{Its \emph{tail coefficient} in case $\td(c)<\infty$ is}
\tc(c)&\:=\liminf_{s\to 0}\frac{c(s\cdot\unitbox{r})}{s^{\td(c)}}
\end{align}
where $s\in(0,1]$. We observe that $\td(c)\geq 1$ for $r\geq 1$. Formally, we set $\tc(c)=0$ if $\td(c)=\infty$ and define the
\emph{tail characteristic} of $c$ as the function
$$
\tchar_c=(\tcoef_c,\tdeg_c)\colon\F\to [0,\infty]\times [0,\infty]\,
$$ 
on the set of front faces $\F$ of $\unitbox{r}$ by 
$\tcoef_c(F)\:=\tc(c_F)$, $\tdeg_c(F)\:=\td(c_F)$.  
Here $c_F$ denotes the push forward measure of $c$ to $F\in\F$ with respect to the 
canonical projection $\unitbox{r}\to F$. Front faces of $\unitbox{r}$ are those faces which
contain the origin. Hence the tail characteristic is the collection of all tail coefficients
and tail degrees of the projections of $c$ to the front faces of $\unitbox{r}$.
We say that a copula $c$ on $\unitbox{r}$ has \emph{tail dependence of degree}~$\td(c)$
if its tail degree satisfies $\td(c)<r$; otherwise it has \emph{no tail dependence}.    
\end{defi}

\begin{exmpl}
The Clayton copula given by 
$\clayton([0,u])=(u_1^{-\theta}+\cdots+u_r^{-\theta}-r+1)^{-1/\theta}$ with parameter 
$\theta\in(0,\infty)$ has 
$
\tchar_{\clayton}(F)=\bigl((\dim F)^{-1/\theta},1\bigr)
$ 
for $F$ with $\dim F\geq 1$. 

The Gumbel copula 
$\gumbel([0,u])=\exp(-((-\ln u_1)^\theta+\cdots+(-\ln u_r)^\theta)^{1/\theta})$, 
for parameter $\theta\in [1,\infty)$ satisfies 
$
\tchar_{\gumbel}(F)=\bigl(1,(\dim F)^{1/\theta}\bigr)
$
for $\dim F\geq 1$. 
Hence the Gumbel copula has tail dependence of degree $r^{1/\theta}$ provided $\theta>1$.  
This must not be confused with the tail dependencies at the opposite vertex $(1,\dots,1)$. 

For convenience and without loss of generality 
we work exclusively with tail dependencies at the origin. 
For $r=2$, lower tail dependence in~\cite{QRM} implies tail dependence of degree~$1$ and
tail coefficient $>0$ for the face $[0,1]^2$. 
We refer to A.~Charpentier~\&~J.~Segers~\cite{CS} who have investigated the tails of
archimedean copulas in a very general setting.  
\end{exmpl}

We consider now a face $F'$ and a face $F$ of $F'$. 
As projecting the copula first to $F'$ and the result to $F$ is the same as projecting 
the copula directly to $F$ we see that $\tdeg$ is a non-decreasing map in the following sense:
If $F,F'$ are two faces of $\unitbox{r}$ and $F\subset F'$, then $\tdeg(F)\leq\tdeg(F')$. 
We call a map $b\colon\F\to\R$ with $b(F)\leq b(F')$ for $F\subset F'$ 
\emph{non-decreasing} and \emph{increasing} if the strict inequality $b(F)<b(F')$ holds for any $F\subset F'$.

\begin{mainres} Let $a,b\colon\F\to (1,\infty)$
denote maps on the faces of $\unitbox{r}$ such that $(a(F),b(F))=(1,\dim F)$ for those 
$F\in\F$ with $\dim F\leq 1$. Let $b$ be non-decreasing.   
\begin{enumerate}
\item
If $b$ is increasing we construct a copula measure $c$ with $\tchar_c=(a,b)$. The copula is locally piecewise uniform in $(0,1)^r$. 
\item
If $b$ is not increasing, we give necessary conditions for $a,b$ such that $\tchar_c=(a,b)$ for some copula~$c$. We investigate special cases where $\tdeg$ is not increasing.   
\item
The construction for the proof generalises naturally to \emph{copulas of order}~$k$, i.e.,   
measures on the unit cube which project to any face of dimension~$k$ to the uniform probability measure.
\item
Tail characteristics for risks $X_1,\dots,X_r$ could be defined via the transformations of the $X_i$ to uniform random variables.  The construction works just as well with  
any other transformation. 
\item
The construction comes along with an efficient simulation algorithm. 
\end{enumerate}
\end{mainres}

Finally we remark that the construction is elementary and contributes to the understanding of dependence patterns for random variables. We can imagine many applications for risk modelling.
Some of them we are going to discuss elsewhere.

The paper is organised as follows. We illustrate and motivate the copula construction principle 
in order to prove the main results by means of two simple examples in the next section. 
In Section~\ref{sec:notation} we introduce some notation and
in Section~\ref{sec:vertexcopulas} we study the spaces of the most simple non-trivial 
copulas in any dimension. They are the building blocks in the construction of our main result. 
We introduce the construction technique of nesting in Section~\ref{sec:nesting}. We explore 
it in Section~\ref{sec:tailnest} for shaping the tail characteristics and eventually   
state Theorem~\ref{thm} about tail nesting. We derive some corollaries in Section~\ref{sec:tailchar} 
and discuss the construction further in Section~\ref{sec:chcoord}.

\begin{danke}
I thank my colleague Guido Gr\"utzner for helpful discussions and the entire Secquaero team for their support. 
\end{danke}

\section{Motivation} \label{sec:motivation}

This section illustrates some of the ideas and observations in this paper 
in a very elementary fashion. 
The reader may find the descriptions in this section helpful when going through 
the construction in any dimension.

\begin{exmpl}[Tail nesting in dimension $2$] \label{exmpl:2d}
We decompose the unit $2$-di\-men\-sion\-al square $[0,1]^2$ into four boxes by splitting 
each edge in the middle, i.e, into $[s,s+1/2]\times [t,t+1/2]$, for $s=0,1/2$ 
and $t=0,1/2$. Each of the four vertices $(i,j)\in\{0,1\}^2$ corresponds to one of these 
squares. Now we choose a probability measure on $[0,1]^2$ which has constant density in
each square. We describe this measure by the map $c'\colon\{0,1\}^2\to [0,1]$ which assigns
to each vertex the measure of the corresponding box in the decomposition. This measure is
a copula if it projects to the uniform measure on each of its edges. In choosing a copula 
$c$ of that type we have only one degree of freedom. We can set the probability
$c'(0,0)=c([0,1/2]^2)$ equal to any $p\in[0,1/2]$. Then, due to the copula condition,
$c'(0,1)=c'(1,0)=-p+1/2=:q$ and thus $c'(1,1)=p$. This is the most simple case of a 
grid copula. The application of grid copulas in risk management was suggested 
by D.~Stra\ss burger \& D.~Pfeifer~\cite{SP}.

\begin{figure}[ht] 
  \begin{center}
    \includegraphics[width=3.5cm]{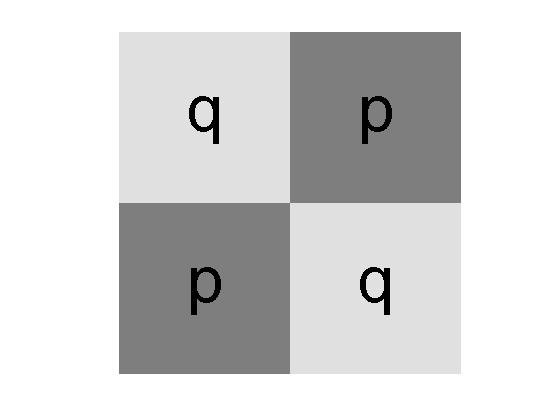} 
    \includegraphics[width=3.5cm]{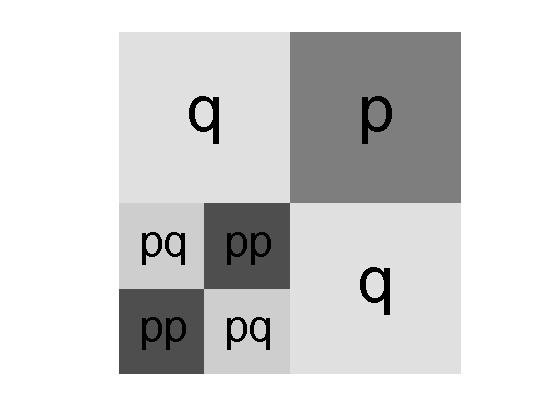}
    \caption{\label{fig:2d} Nesting a $2$-dimensional box copula into itself. Grey levels 
             according to the density of probability for some $p\in(1/4,1/2)$.}
 \end{center}
\end{figure} 

The following observation is simple, but crucial for the remainder of the paper: 
We set $c^1=c$ as above.  Then we \emph{nest} $c$ into the square 
$[0,2^{-1}]^2$ of $c^1$ as follows: Decompose that square again into four equally 
sized squares $[s,s+2^{-2}]\times [t,t+2^{-2}]$, for $s=0,2^{-2}$ and $t=0,2^{-2}$ and 
modify $c^1$ on $[0,2^{-1}]^2$ by `multiplying' it with $c$, in order to obtain 
$c^2$, as illustrated in Fig.~\ref{fig:2d}.

In this construction we have refined the initial decomposition of $[0,1]^2$. We call such 
decompositions \emph{box decompositions}. It can be verified immediately that $c^2$ is
again a copula. We call copulas of that type \emph{box copulas}. Now we can repeat this 
construction by nesting $c$ into the square $[0,2^{-2}]^2$ of $c^2$, in order 
to obtain $c^3$ and so on, by recursively nesting $c$ into the square 
$[0,2^{-n}]^2$ of $c^n$. The limit $c^\infty$ of this sequence of copulas exists and 
is again a copula. Suppose we start with $p=(1/2)^b$ 
and $b>1$. Then $c^\infty([0,2^{-n}]^2)=p^n=(2^{-nb})$ and thus 
$\limsup  c^\infty([0,u]^2)/u=0$ as $u\to 0$. 
\end{exmpl}

\begin{rem}
Key observations when studying the simple example are:
\begin{enumerate}
	\item The copulas $c^n$ are asymmetric and the probability density increases as 
	      $u\to 0$ for $p>1/4$.
	\item As $(u/2)^b < c^\infty([0,u]^2)\leq u^b$,the copula $c^\infty$ has zero
	      lower tail dependence\footnote{$c^\infty$ is said to have lower tail dependence if 
	      $\lim c^\infty([0,u]^2)/u>0$ as $u\to 0$, refer e.g.\ to \cite{QRM} for  details.} 
	      for $b>1$. 
	\item Nevertheless, given $p\in [0,2^{-n}]$, we can choose $b$ such that 
	      $c^\infty([0,2^{-n}])=p$. Hence this copula family is still good enough for
	      sensitivity testing in risk modelling. Furthermore, 
	      there is a simple recursive algorithm to generate samples of $c^\infty$.    	 
	\item We can further modify the tail behaviour by nesting in the $n$-th step a copula 
	      of the same type but with $c'(0,0)=(1+\delta_n)\cdot 2^{-b}$ where $|\delta_n|$ is close to~$0$. 
	      By choosing appropriate sequences $(1+\delta_n,b)$ we can not only control the 
	      tail dependencies in the limit but also how the limit is approached 
	      as $n\to\infty$.
\end{enumerate}
\end{rem}

\begin{exmpl}\label{exmpl:frac}
Analogously to the decomposition in Example~\ref{exmpl:2d}, we decompose now the unit 
$3$-dimensional cube $[0,1]^3$  into $8$ cubes, each isometric to $[0,1/2]^3$. Each of these 
cubes contains exactly one of the vertices $\nu\in\{0,1\}^3$. We assign to the `even' 
cubes (i.e., those with $\nu_1+\nu_2+\nu_3\equiv 0 \mod 2$) the uniform measure with 
total probability equal to $1/4$. The `odd' cubes get probability zero. 
When projected to any $2$-dimensional face, the resulting probability measure 
is the uniform probability measure. In particular, $c$ is a copula. 


\begin{figure}[ht] 
  \begin{center}
    \includegraphics[width=3.2cm]{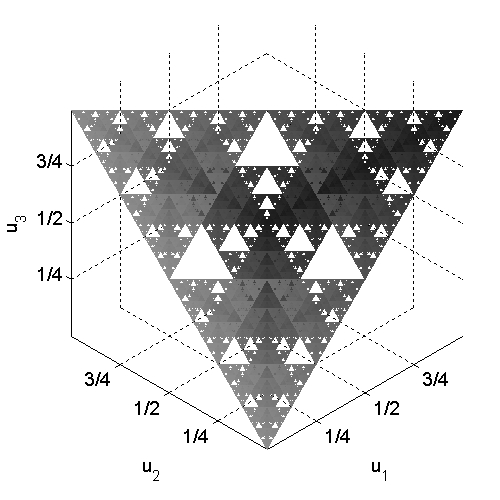}
    \includegraphics[width=3.2cm]{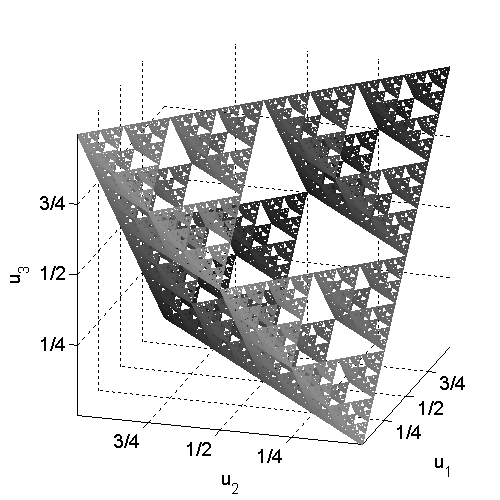}
    \includegraphics[width=3.2cm]{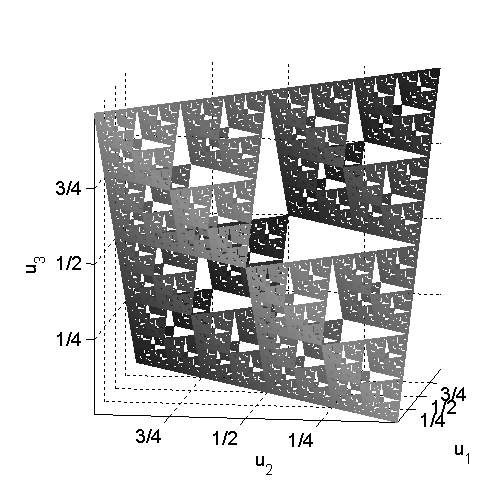}
    \includegraphics[width=3.2cm]{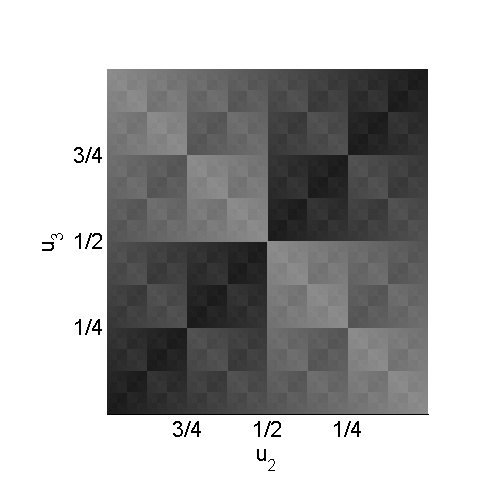}
    \caption{\label{fig:frac} Illustration of $c^9$ from Example~\ref{exmpl:frac}. 
             The pictures were generated by plotting one point in the centre of each of the $4^9$ 
             `even' cubes. The grey level of a point $(u_1,u_2,u_3)$ is given by $-u_1$. 
             The view in the left picture is along a diagonal. From left to right 
             it is stepwise rotated around the third coordinate axis. The points in the 
             picture on the right are uniformly distributed in the $(u_2,u_3)$-plane. 
             Their grey level is merely an indication of the $u_1$-level.}
 \end{center}
\end{figure}

Now set $c^1\:=c$ and nest $c$ into the even cubes of the
decomposition underlying $c^1$. In this way the even cubes decompose again into $8$ cubes,
each isometric to $[0,1/4]^2$; four are again `even' and the others are odd. 
We obtain a copula $c^2$ on $[0,1]^3$, its support consisting of $16$ cubes, 
each with uniform measure and probability equal to $(1/4)^2$. Observe now that $c^2$ 
still projects to the uniform measure on each of the $2$-dimensional faces of $[0,1]^3$. 
We can continue this nesting and obtain a limit measure $c^\infty$. 

The projection of the limit measure to any $2$-face of $[0,1]^3$ is again the uniform measure 
and thus $c^\infty$ is in particular a copula. It is not difficult to see that 
the limit measure is, up to scaling, the $2$-dimensional Hausdorff measure of the support 
of $c^\infty$. The latter is the intersection of all the `even' cubes obtained during the 
recursive definitions of the  $c^n$. 
\end{exmpl}

\begin{rem} We summarise the main observations from the above example. To this end assume 
that we have three risks $X_1,X_2,X_3$ whose dependence structure is given by $c^\infty$, 
i.e., $P(X_i\leq Q_i(u_i),i=1,2,3)=c^\infty([0,u])$ where $Q_i$ is the quantile function. 
\begin{enumerate}
	\item Even if risks $X_1,X_2,X_3$ are pairwise independent, they can be heavily dependent overall. 
	      For the univariate margins of the above copula the third margin is a function of 
	      the other two. 
	\item The probability that all three risks are worse than their $2^{-n}$-Quantile is
        $\bigl(2^{-n}\bigr)^2$.
	      A measure for tail dependencies in higher dimensions should show that $c^\infty$
	      has some tail dependence. This is one motivation for Definition~\ref{defi:tdep}.
\end{enumerate}
\end{rem}

It has become more and more common that risk modellers focus on tail dependencies when modelling 
a portfolio. This example demonstrates nicely that in calibrating the corresponding dependence 
models it is not sufficient to focus on the estimation of pairwise dependencies alone.

The construction for shaping the tails of copulas in Section~\ref{sec:tailnest} is as in
Example~\ref{exmpl:2d}, but generalised to any dimension $r\geq 2$. 
Roughly speaking, we can shape the projections to any lower dimensional faces, 
which are also of this type, simultaneously so that we can achieve any tail characteristic
which is consistent with the condition for probability measures.  

Before describing this aspect, we need to define some notation related to cubes, their vertices 
and faces in Section~\ref{sec:notation}.  Then we study maps $\{0,1\}^r\to[0,1]$, the 
equivalents of those maps for $r=2$ in Example~\ref{exmpl:2d}, 
 which define copulas in Section~\ref{sec:vertexcopulas}.

\section{Notation and basic definitions} \label{sec:notation}

%
%

By $u,v,w$ we denote usually points in $\R^r$ where $u=(u_1,\dots,u_r)$.
The unit $r$-\emph{cube} is $\unitbox{r}\subset\R^r$. The set of its 
\emph{vertices} is $$ V:=\{0,1\}^r\,.
$$
We use the letters $\nu,\mu$ exclusively for elements of $V$. 
There is a one-to-one correspondence between \emph{front faces} 
$F$ of the $r$-cube and the vertices $\nu\in\{0,1\}^r$ given by 
\begin{equation}
F(\nu) =\dset{u\in\unitbox{r}}{$u_i=0$ if $\nu_i=0$}
\end{equation}
We set $\zero:=(0,\dots,0)$ and $\one:=(1,\dots,1)$.  
and denote by 
$$
\,\widehat{\,\,\,}\colon u\mapsto\widehat u=\one-u
$$ 
the reflection with $\hat\zero=\one$. 
For a front face $F$ the corresponding back face is $\widehat F$ and 
the complementary front face $F^c$. We observe that the front face complementary 
to $F(\nu)$ is $F(\hat\nu)$.
We can identify each face $F$ naturally with $\unitbox{\dim F}$. 
Other faces of $\unitbox{r}$ are of the form $F(\nu)+\mu$ for $\mu\in F(\hat\nu)$.
Given a front face $F$ of $\unitbox{r}$ we denote by 
\begin{equation}
\pi^F\colon\unitbox{r}\to F^c
\label{eq:faceproj}
\end{equation}
the \emph{canonical projection along} $F$ to its complement $F^c$. 

%
%
An interval $I$ in $\R^r$ is an $r$-fold product of intervals in $\R$.  We call a 
compact interval in $\R^r$ with non-empty interior an $r$-\emph{box} in $\R^r$. 
Hence $r$-boxes are of the form $[u_1,v_1]\times\cdots\times [u_r,v_r]=:[u,v]$ with $u_i<v_i$ 
for each $i$. A \emph{box decomposition} $\I$ of an $r$-box $I$ is a collection of finite $r$-boxes 
$\{\I_1,\dots,\I_n\}$ such that their union is $I$ and their non-trivial intersections are of 
lower dimensions, i.e., $I_1\cup\cdots\cup I_n =I$ and $\interior{I_k}\cap\interior{I_l}\neq \emptyset$
for all $I_k\neq I_l$.

Given an $r$- and $r'$-box $I$ and $I'$ with box decompositions $\I$ and $\I'$, respectively, 
we can form the product box $I\times I'$ in $\R^{r+r'}$ which inherits a natural box decomposition 
$\I\times\I'$, the product decomposition. If a box decomposition of a box $I$ is the product of box 
decompositions of its $1$-dimensional faces we call it a \emph{grid decomposition} 
or simply a \emph{grid}.

Given an $r$-box $I=[u,v]$ we denote by $\imath$ the canonical affine transformation from 
the unit $r$-cube $\unitbox{r}$ to $I$,
\begin{equation}
\imath\colon\unitbox{r}\to I,\quad, w\mapsto \big(u_1+w_1(v_1-u_1),\dots,u_r+w_r(v_r-u_r)\big)
\end{equation}
Via this transformation we define the corresponding faces of and projections to faces of $I$. 
Observe that $\imath$ maps box decompositions $\I$ of $\unitbox{r}$ to box decompositions of $I$.

\begin{defi}[Box measure]
Let $\I=\{I_1,\dots, I_n\}$ be a box decomposition of $I$ and $\unitbox{r}$. 
We view elements $b\in\R^\I$, i.e., maps $b\colon\I\to\R$, as signed measures on $I$ such that
$b|_{I_i}$ is a uniform measure, i.e., proportional to the Lebesgue measure of $I_i$. 
\end{defi}

Let $\J$ be a box decomposition of $\unitbox{r}$. The canonical transformations 
$\imath\colon\unitbox{r}\to I$ induces a canonical vector space isomorphism
$\imath\colon \R^\J\to\R^{\imath(\J)}$ by pushing forward the measures. 

In the same way we can push forward box measures using the 
canonical projections $\pi^F$ along faces.
In case $\J$ is not a grid decomposition, the elements of $\pi^F(\J)$ do not define a 
box decomposition of $F^c$ (see e.g., Fig.~\ref{fig:2d}). 
In this case, we denote by $\pi^F(\J)$ the projection of \emph{some} refinement 
of $\J$ into a grid. In this way, we can push forward box measures in $\R^\J$ to
box measures on $F^c$ and obtain a linear map 
\begin{equation} \label{eq:piFforward}
\pi^F\colon\R^\J\to\R^{\pi^F(\J)}\,.
\end{equation}
Our ultimate goal is to construct copulas in $\R^\J$ satisfying some given conditions, 
such as a certain behaviour near the the origin $\zero\in\unitbox{r}$. 
To this end the following notion turns out to be quite useful.

\begin{defi}
Let $\J$ be a box decomposition of $\unitbox{r}$. 
Then define the following linear subspaces of $\R^\J$ for $k=0,1,\dots,r$, 
\begin{equation} \label{eq:Gk}
G_k(\J)=\bigcap\Dset{\ker\bigl(\pi^F\colon\R^\J\to\R^{\pi^F(\J)}\bigr)}{$F\in\F$, $\codim(F)=k$}
\end{equation}
and set $G_{-1}(\J)=\R^\J$.  
\end{defi}
Observe that $G_k(\J)$ is the set of those box measures which project onto each face
of dimension $k$ to the zero measure. Thus it does not depend on the choice
of $\pi^F(\J)$ in~\eqref{eq:piFforward}. Furthermore, we obtain a filtration of 
linear subspaces of $\R^\J$,
\begin{equation} \label{eq:filtrG}
\{0\}=G_r(\J)\subset G_{r-1}(\J)\subset\cdots
\subset G_0(\J)\subset G_{-1}(\J)=\R^\J
\end{equation} 
We let $\unifc_\J\in\R^\J$ denote the element which corresponds to the 
uniform probability measure on $\unitbox{r}$.

\begin{defi}For $k=0,1,2,\dots,r$ we define the following subsets of $\R^\J$:
\begin{equation}
C_k(\J)=\bigl(\unifc_\J+G_k(\J)\bigr)\cap [0,1]^\J\,.
\end{equation}
The set of probability measures in $\R^\J$ is $C_0(\J)\subset\R^\J$ and 
the subset of copulas is $C_1(\J)$, the \emph{box copulas} in $\R^\J$.
We call elements of $C_k(\J)$ \emph{box copulas of order}~$k$. We observe that 
$$
\pi^F\colon C_k(\J) \to C_k(\pi^F(\J))
$$ 
for $k\leq\codim F$. The sets $C_k(\J)$ are convex in $\R^\J$. We call elements of
$G_k({\J})$ \emph{copula generators of order} $k$ for $\J$.      	 

The limits of sequences of box copulas we work with later on are not 
box copulas. We extend some notions from above to Borel measures 
on $\unitbox{r}$ such as the property of being a 
\emph{copula of order}~$k$ in the obvious way.
We call a probability measure $c$ on $\unitbox{r}$ 
\emph{locally piecewise uniform on} $E\subset\unitbox{r}$ if for each
$w\in E$ there exists a neigbourhood $W$ of $w$ in $\unitbox{r}$ such that
$c|_W$ is equal to some box measure restricted to $W$.  
\end{defi}

\begin{rem}
Note that by construction each box copula corresponds to a probability measure
on $\unitbox{r}$ such that the projection to any edge yields the uniform
probability measure. The notion of a box copula differs from the notion 
of a \emph{grid-type copula}~\cite{SP} or simply \emph{grid copula}
only by the underlying decomposition. 
We already made use of the fact that any box decomposition can be 
refined into a grid decomposition when defining projections.  
One reason for introducing the notion of a box copula is that it is 
useful when designing efficient simulation algorithms for nested copulas
(see Algorithm~\ref{algo:tailnest}). For that purpose, we may want to describe
the measure with the smallest possible number of boxes
(see Fig.~\ref{fig:2d} for an illustration).  
\end{rem}

\begin{exmpl}[Box copulas] It is not difficult to construct box copulas:
\begin{enumerate}
\item  The copulas $c^n$, $n<\infty$, from Example~\ref{exmpl:frac}
       are box copulas of order~$2$. 
\item  Take any copula $c'$ on $\unitbox{r}$, select some $\J$ and 
       let $c(J)$ be the $c'$-measure of $J$, $J\in\J$. 
       Then $c\in C_1(\J)$. The finer $\J$, the better the 
       approximation of $c'$ by $c$.    
\item  Given some copula generator $g\in G_k(\J)$, some 
       $c\in C_k(\J)\cap (0,1)^\J$, then $c+tg\in C_k(\J)$ 
       for $|t|\leq\epsilon$ and $\epsilon>0$ sufficiently small. 
       This follows from the convexity of $C_k(\J)$. 
\end{enumerate}
\end{exmpl}

\begin{rem}[Copula surgery] 
If $I\subset\unitbox{r}$ is an $r$-box, $g\in G_k(\J)$ and 
$\imath\colon\unitbox{r}\to I$ the canonical map, then 
$\imath(g)$ can be viewed as a copula generator on $\unitbox{r}$ with support 
contained in~$I$ and with respect to any box decomposition $\I$ which extends $\imath(\J)$. 
Observe that we can now build the sum $c+\imath(g)$ as signed measures for $c\in C_k(\I')$. 
Provided $g$ is appropriate, e.g., $|g(\nu)|$ sufficiently small for each $\nu$, 
$c+\imath(g)$ is a copula of order~$k$. 
It is indeed a box copula for any common refinement of $\I'$ and $\I$. 
This construction can be iterated with appropriate sequences of generators. 
The reader may wish to compare this with some of the methods
for the construction of copulas in~\cite{Nel}.

Nesting and tail nesting which we introduce later on are special cases of `copula surgery'. 
In order to explore the range of such constructions, 
a detailed study of the vector spaces $G_k(\J)$ for certain $\J$ turns out to be helpful. 
This is the subject of the next section. 
\end{rem}

\section{Vertex decompositions and their copulas} \label{sec:vertexcopulas}

In this section we will describe the box copula spaces for the most 
simple box decompositions. A point $u\in(0,1)^r$ induces a decomposition
of the $i$-th edge, namely $\{[0,u_i],[u_i,1]\}$. The product of these 
decompositions is a box decomposition $\J(u)$. Each box of this 
decomposition contains exactly one vertex of $\unitbox{r}$. In this way, 
we identify 
$$\J(u)\simeq V=\{0,1\}^r$$ 
via~$u$. We call these decompositions \emph{vertex decompositions} and corresponding
box measures \emph{vertex measures}. Vertex measures are defined by $u$ together
with a map $\{0,1\}^r\to \R$, i.e, by an element of $\R^V$. 
We denote the canonical basis of $\R^V$ by $(e_\nu)_{\nu\in V}$.
Observe that the linear spaces $G_k(V)$ do not depend on $u$; but the copula spaces
$$
C_k(u):=C_k(\J(u))
$$
do. 
We also point out that vertex decompositions project to 
vertex decompositions along the faces $F$ via the corresponding map
$\pi^F\colon\R^V\to\R^{V\cap F^c}$.  

Given a subset $W\subset V$ we view $\R^W$ naturally as the linear subspace of $\R^V$.
In this way a map $W\to\R$ is extended to a map on $V$ by mapping 
elements in $V\setminus W$ to zero. The canonical projection $\R^V\to\R^W$ 
corresponds to restricting maps $V\to\R$ onto $W$, denoted by $x\mapsto x|_W$.
We decompose $V$ into disjoint sets
\begin{align} 
V&=V_r\cup V_{r-1}\cup\cdots\cup V_0
\label{eq:Vr}
\intertext{with $V_k=\tset{\nu\in V}{$\nu_1+\cdots+\nu_r=r-l$}$.  We write}
V_{\geq k}&=V_r\cup V_{r-1}\cup\cdots\cup V_k\,. \label{eq:Vdecompr0}
\intertext{and likewise $V_{>k}$ or $V_{\leq k}$ and so forth. Accordingly, we decompose the set of 
front faces $\F$ of $\unitbox{r}$ into}
\F&=\F_r\cup\F_{r-1}\cup\cdots\cup\F_0
\end{align}
where $\F_l$ is the \emph{set of front faces of codimension}~$l$. 
Observe that $V_r=\{\zero\}$ and $\F_r=\{\{\zero\}\}$. We denote by
$ 
\F^k:=\F_{r-k}
$
the set of \emph{front faces of dimension}~$k$.  We also write $\F_{>k},\F^{\leq k},\dots$ along the lines of~\eqref{eq:Vdecompr0}. 

\newpage

\begin{rem} We summarise some observations for later purposes\footnote{Drawing pictures for 
$r=3$ may provide helpful illustrations.}.
\begin{enumerate}
	\item The map map $V\to\F$, given by $\nu\mapsto F(\nu)$  is compatible with the 
	      above decompositions. It defines bijections $V_k\to\F_k$, $k=0,1,\dots,r$. 
	      For $\nu\in V_k$ we have that $F(\nu)\subset V_{\geq k}$ and $F\cap V_k=\{\nu\}$.
	      In the same way, the map $\nu\mapsto F(\hat\nu)$  defines bijections $V_k\to\F^k$. 
	\item Choose $x\in\R^V$ and $\nu\in V$. Then $\pi^{F(\nu)}(x)\in\R^{F(\hat\nu)\cap V}$ with 
	      $$\pi^{F(\nu)}(x)(\mu)=\sum_{\nu'\in F(\nu)+\mu} x(\nu')\,.$$
  \item $G_k(F\cap V)\subset G_k(V)$ for each $F\in\F$ and any $k$ where $G_k(F\cap V)$
        is defined. Indeed, given $x\in G_k(F\cap V)$ and $F'\in\F_k$ 
        we observe that $F'\cap F$ has codimension $\leq k$ in $F$. Hence 
        $x\in G_k(F\cap V)\subset\R^V$ projects to zero along $F'$. 
  \item We have that $\pi^F(G_k(V))\subset G_k(F^c\cap V)$. Indeed, if $F'$ is a front 
        face of codimension $k$ in $F^c$, then $F\times F'$ is a face of 
        codimension $k$ in $\unitbox{r}$. Hence $\pi^{F'}(\pi^F(x))=\pi^{F\times F'}(x)=0$
        for $x\in G_k(V)$.
\end{enumerate}
\end{rem}

Next we are describing the spaces $G_k(V)$ and the corresponding copula spaces.

\begin{lem} \label{lem:GkvpiF}
The vector spaces $G_k(V)\subset\R^V$ have the following properties:
\begin{equation}
G_k(V)=\dset{x\in\R^V}{$\pi^F(x)(\zero)=0$ for each $F\in\F_{\leq k}$}
\label{eq:GkpiF0}
\end{equation}
In other words, $x$ projects to the zero-measure along any face of codimension $k$ 
if and only if the projection of $x$ along any face of codimension 
$\leq k$ is zero at the origin $\zero$.  
\end{lem}

\begin{proof}[Proof of Lemma~\ref{lem:GkvpiF}]
It is evident that $G_k(V)$ is contained in the set on the right hand side of~\eqref{eq:GkpiF0} 
which we denote by $G'_k(V)$. We argue by induction over the dimension $r$. To this end we note 
that the statement is trivial for $r=0$. We assume that 
$G'_k(V')=G_k(V')$ for the vertices $V'$ of cubes up to dimension $r-1$ and for all $k=0,\dots,r-1$. 
Suppose now that $x\in G'_k(V)$ and $F\in\F_k$. We need to show that $\pi^F(x)=0$. 
We distinguish two cases. 

Firstly, assume $k<r$.  Then we observe that 
$\pi^F(x)\in G'_k(F^c\cap V)$ and by the induction hypothesis in $G_k(F^c\cap V)$ as 
$\dim F^c=k<r$. Note that $G_k(F^c\cap V)=\{0\}$ as $\dim F^c=k$ and hence $\pi^F(x)=0$. 

Secondly, If $k=r$ we see from the definition of $G_r(V)$ that $x$ restricted to any 
$F\in\F_1$ is in $G'_{r-1}(F\cap V)$ and thus in $G_{r-1}(F\cap V)$. 
Hence $x(\nu)=0$ for $\nu\in \bigcup \F_1=V\setminus\{\one\}$. 
It remains to show that $x(\one)=0$. This follows by projecting
$x$ along the $r$-dimensional front face $\unitbox{r}$ to the zero-dimensional face $\{\zero\}$ 
which yields $0$ as $x\in G'_r(V)$ by assumption. Hence $x(\one)=0$.
\end{proof}

The following proposition describes the degrees of freedom one has
in finding box copulas with prescribed properties for the tail.

\begin{prop} \label{prop:GkV}
There is a linear map $S\colon\R^V\to\R^V$ with the following properties: 
\begin{enumerate}
	\item $S$ is an isomorphism between the vector space filtrations, i.e., 
	      the following diagram is commutative for $S$:
\begin{align*} 
\{0\}&=&\R^{V_{>r}}&\subset&\R^{V_{>r-1}}&\subset\cdots\subset&\R^{V_{>0}}&\subset &\R^{V_{>-1}}&=\R^V\\
     &&\downarrow^{_\simeq}\ \ &&\downarrow^{_\simeq}\ \ &&\downarrow^{_\simeq} \ \ &&\downarrow^{_\simeq} \ \ & \\
\{0\}&=&G_r(V)&\subset & G_{r-1}(V)&\subset\cdots
\subset &G_0(V)&\subset &G_{-1}(V)&=\R^V 
\end{align*}
\item For $x\in\R^V$ and any $F\in\F$ we have $\pi^{F(\nu)}(S(x))(\zero)=x(\nu)$.
\end{enumerate}
$S$ is uniquely determined by~(i) and~(ii) and is given by
\begin{equation} \label{eq:Sxmu}
S(x)=\sum_\nu\sum_{\mu\in F(\nu)}(-1)^{\mu-\nu}x(\mu)e_\nu
\end{equation} 
for $x\in\R^V$ where $(-1)^\nu\:=\prod_i(-1)^{\nu_i}$ for any $\nu\in V$.
\end{prop}
Eventually we are interested in the construction of copulas with prescribed tail properties. 
The proposition will be instrumental for this.

\begin{rem}
Observe that the inverse of $S\colon\R^V\to\R^V$ is given by
\begin{equation}
S^{-1}(z)(\nu)=\pi^{F(\nu)}(z)(\zero)=\sum_{\mu\in F(\nu)}\!z(\mu)\,,\quad \nu\in V\,. 
\end{equation}
\end{rem}

\begin{proof}[Proof of Proposition~\ref{prop:GkV}]
If $S$ with properties~(i) and~(ii) existed, it would be unique as its inverse would be given by the previous remark. To prove existence, we just construct~$S$.
First we are going to define $S'(e_\zero):=\sum_\nu(-1)^\nu e _\nu$ and it 
is readily verified that $S'(e_\nu)\in G_{r-1}(V)$. 
From the previous lemma we know that $\dim G_{r-1}(V)=1$.
For any $\mu\in V_k$ we denote by 
$$
\widehat{F}(\mu):=\widehat{F(\hat\mu)}
$$ 
the back face corresponding to $F(\hat\mu)$. 
Observe that $\dim\widehat{F}(\mu)=k$ and that $\mu$ corresponds naturally to 
the origin in $\widehat{F}(\mu)$.  Exactly as with $S'(e_\zero)$ before, 
we define now $S'(e_\mu)$ to be the unique element in 
$G_{k-1}(\widehat{F}(\mu)\cap V)\subset G_{k-1}(V)$ having the required property, 
and that is
\begin{align}
S'(e_\mu)& =\sum_{\nu\in\widehat{F}(\mu)}(-1)^{\mu-\nu}e_\nu\,. \\
\intertext{Therefore we obtain a candidate for the isomorphism in question, namely} 
S'(x)&= \sum_\mu\sum_{\nu\in\widehat{F}(\mu)}(-1)^{\mu-\nu}x(\mu)e_\nu\,.
\end{align}
Now observe that $\nu\in\widehat{F}(\mu)$ if and only if $\mu\in F(\nu)$ and thus 
$S'=S$ where $S$ is as in~\eqref{eq:Sxmu}. As $S$ is compatible with the filtration 
on the basis $(e_\mu)$, it is indeed compatible with the filtration. As it satisfies
$\pi^{F(\nu)}(S(e_\mu))(\zero)=1$ for $\nu=\mu$ and $\pi^{F(\nu)}(S(e_\mu))(\zero)=0$ 
otherwise it satisfies $\pi^{F(\nu)}(S(x))(\zero)=x(\nu)$ by linearity. 
By Lemma~\ref{lem:GkvpiF}, $S$ restricts to isomorphisms $\R^{V_{>k}}\xrightarrow{_\simeq}G_k(V)$.
\end{proof}

\begin{rem} \label{rem:Tk} 
For each $k$ there exists a unique map 
$T_k\colon\R^{V_{>k}}\to G_k(V)$ such that $T_k(x)(\nu)=x(\nu)$ for $\nu\in V_{>k}$.
The map $T_k$ is a vector space isomorphism.
As we do not make use of this result, we leave the proof as an exercise.
\end{rem}


Next we describe the vertex copulas for $\J(u)$.  We abbreviate 
$$
u^\nu:=\prod_{\{i\,|\,\nu_i=1\}} u_i
$$ for any $\nu\in V$ and $u\in(0,1)$ where $u^\zero=1$. 
The uniform copula in $C_k(u)$ corresponds to the element 
$$
\unifc_u:=\sum_{\nu\in V} u^{\one-\nu}(\one-u)^\nu e_\nu
\in\R^V
$$
which depends on $u$. Observe that $\pi^{F(\nu)}(\unifc_u)(\zero)=u^{\one-\nu}$. 
Therefore the proposition implies that
$$
S(x_u)=\unifc_u\quad\text{for\ }x_u=\sum_\nu u^{\one-\nu} e_\nu\,.
$$ 
\begin{cor} \label{cor:Ck}
The vertex copulas of order $k$ with respect to $\J(u)$ are given by
$$
C_k(u)=\dset{S(x)}{$x\in\R^V$, $x(\nu)=u^{\one-\nu}$ for $\nu\in V_{\leq k}$, 
$x$ satisfies~\eqref{eq:boxcopcond}}
$$
with
\begin{equation}
\sum_{\mu\in F(\nu)}(-1)^{\mu-\nu}x(\mu)\in[0,1]\quad\text{for each\ } \nu\in V	\,.
	\label{eq:boxcopcond}
\end{equation}
Condition~\eqref{eq:boxcopcond} is the condition for $x$ being a probability measure. 
\end{cor}

\begin{proof}
A given element in $c\in C_k(u)$ is of the form $c=g+\unifc_u$ with $g\in G_k(V)$. 
Using the proposition, choose $y\in\R^{V_{>k}}$ such that $S(y)=g$ and set $x=y+x_u$.
Then $S(x)=c$ and the statement follows from the proposition. 
\end{proof}


\begin{rem}
The reader may compare Condition~\eqref{eq:boxcopcond} with the rectangle inequality 
for a copula in~\cite{QRM}, p.~185. We see from the corollary that in case of vertex measures
there is no need to check that property for every $r$-box in $\unitbox{r}$.
\end{rem}

\section{Nesting box copulas} \label{sec:nesting}

In this section we define formally the nesting constructions used in the
Examples~\ref{exmpl:2d} and~\ref{exmpl:frac}. 
Related techniques have been used by G.~Fredricks et al.~\cite{FNL} 
in dimension~$2$.

\begin{defi}[Nesting box copulas]
Suppose $z_j=\in\R^{\I_j}$, $j=1,2$ are two box measures on $\unitbox{r}$ and 
$I_1\in\I_1$ is a box for $z_1$. Let $\imath\colon\unitbox{r}\to I_1$ be 
the canonical transformation. Then $\nest(z_2,z_1,I_1)$ is the box measure 
$y\in\R^\J$ with $\J=(\I_1\setminus I_1)\cup\imath(\I_2)$ and
\begin{equation}
 y|_{\I_1\setminus I_1}=z_1|_{\I_1\setminus I_1}\quad\text{and}\quad
 y|_{I_1}              =z_1(I_1)\imath(z_2)\,.
\end{equation}
In other words, $\nest(z_2,z_1,I_1)$ is equal to $z_1$ outside of $I_1$ and 
on $I_1$, it is a copy of $z_2$, scaled by $z_1(I_1)$. 
We call $\nest(z_2,z_1,I_1)$ the \emph{box measure obtained from nesting 
$z_2$ into $z_1$ along} $I_1$.  

If $\Scal_1\subset\I_1$, we define $\nest(z_2,z_1,\Scal_1)$ to be the box measure 
obtained by consecutively nesting $z_2$ into the elements of $\Scal_1$. 
Note that this construction does not depend on the order of the different 
nesting operations. We call $\nest(z_2,z_1,\Scal_1)$ 
the box measure obtained from nesting $z_2$ into $z_1$ along $\Scal_1$.  
\end{defi}

\begin{lem} \label{lem:Ckinvundernest}
Assume that $z_j\in C_k(\I_j)$, $j=1,2$ are box copulas of order $k$. 
Then $\nest(z_2,z_1,I_1)$ is in $C_k(\J)$ where $\J$ is as in the definition. 
\end{lem}

\begin{proof}
The measure $z_1|_{I_1}$ is 
uniform.
The projections of $z_1(I_1)\imath(z_2)$ along faces of $I_1$ up to codimension~$k$ 
is by assumption a constant times the euclidean volume of the face for the 
corresponding dimension. The scale factor is arranged such that projections  
of $z_1|_{I_1}$ to these faces are equal to the projection of $z_1(I_1)\imath(z_2)$.
To this end note that $\imath(z_2)(I_1)=1.$
\end{proof}

\begin{algo}[for sampling nested box copulas]  \label{algo:nest}
Suppose that we have a simulation algorithm for drawing random samples 
from copulas $z_1$ and $z_2$ as in the lemma, i.e., essentially an algorithm for 
simulating multinomial distributed variates. Then the following is a 
simulation algorithm for $\nest(z_2,z_1,I_1)$: 
\begin{enumerate}
	\item Draw a random variate $U\in[0,1]^r$ from $z_1$.
	\item If $U\not\in I_1$ then return $U$, 
	\item otherwise generate a random sample $V$ from $z_2$. Return $\imath(V)$. 
\end{enumerate}
\end{algo}

We can nest copulas iteratively. To this end we start with a box copula $c^1$. 
Then construct $c^n$ recursively. Suppose we have already constructed 
$c^{n-1}\in\R^{\J_{n-1}}$. Then we choose a box copula $z_n$, a subset 
$\Scal_{n-1}\subset\J_{n-1}$ and set $c^n=\nest(z_{n-1},c^{n-1},\Scal_{n-1})$.

\begin{rem} Suppose that $c^n\in\R^{\J_n}$, $n>1$,  is a obtained from 
consecutive nestings. Intuitively, the sequence of measures $(c^n)$ 
converges as for $J\in\J_n$, we have that $c^m(J)=c^n(J)$ for each $m\geq n$. 
Formally, given $E\subset[0,1]^r$ define
\begin{align*}
c^\infty(E)\:=\lim_{n\to\infty}\inf\Bigl\{\,\sum_{i}c^{m(i)}(W_i) \,\,\Big| \  &  (W_i)  \text{ a countable open covering of } E, \\
                                                         &  m(i)\geq n\,\Bigr\} 
\end{align*}
Arguing as in \cite{GMT}, p.~8, Eq.~(1), we conclude that $c^\infty$ 
is an outer measure where Borel sets are measurable. 
Form the construction we see that $c^\infty(J)=c^n(J)$ for each $J\in\J^n$. 
It is evident that $c^\infty$ is a copula measure. 
Note that in the interesting case where we have infinitely many non-trivial 
nestings, the limit measure is not given by a box copula.   
\end{rem}

\begin{exmpl}[generalising Example~\ref{exmpl:frac}] \label{exmpl:hdfflimit}
We choose the vertex decomposition of $\unitbox{r}$ given by $u=(1/2,\dots,1/2)$ and  
start with the vertex copula $c^1=2^{-r}S(e_\zero)+\unifc\in C_r(\J(u))$ on $[0,1]^r$,
$S$ as in Proposition~\ref{prop:GkV}.  This copula assigns to each vertex 
$v\in\{0,1\}^r$ with $\mod(\sum{v_i},2)=0$ the probability $2^{-r+1}$ and to the other 
vertices the probability $0$. Proceeding as above, we choose $z_n=c^1$ and 
$\Scal_{n-1}=\J_{n-1}$. The limit copula $c^\infty$ is, up to scaling, 
equal to the $(r-1)$-dimensional Hausdorff measure on the limiting 
set $\bigcap_{n}\support(c^n)$. 

We consider now the canonical projections $U_i\colon\support(c^\infty)\to [0,1]$ 
onto the $i$-th coordinate and view it as a random variable 
with probability measure given by $c^\infty$. We write
$$
U_i=\sum_{n=1}^\infty B_{in}\cdot 2^{-n}
$$
with $B_{in}$ having values in $\{0,1\}\simeq\Z/2\Z$. They are the digits for the 
binary representation of $U_i$. 
It can be seen from the construction of $c^\infty$ that 
$B_{kn}=\sum_{i\neq k} B_{in}$ where the summation is in $\Z/2\Z$. Hence $U_k$ is the 
bitwise addition of the $U_i,i\neq k$. Any subset of $\{U_1,\dots,U_r\}$ 
with less than~$r$ elements is independent.  

This example is well known, at least for finitely many digits 
(where the $U_i$ are `cut off' after the $m$-th digit), 
refer e.g., to J.~Stoyanov~\cite{Stoyanov}. The set of 
$U_i$ can be enlarged by $U_N\:=\sum_{i\in N}^{2}U_i$ for any non-empty subset 
$N\subset\{1,\dots,r-1\}$ where $\sum^2$ stands for bitwise addition as 
described before. In this way we obtain a longer finite sequence of 
pairwise independent random variables $U_N$.  

A similar construction works also for representations with respect to any base $d\geq 2$, 
i.e., with any $\Z/d\Z$. Just start with the 
regular grid for $\unitbox{r}$ which decomposes each edge into $d$ intervals 
of length $1/d$ and assign to them the digits $0,1,\dots,d-1$.
Define the starting box copula as above for the case $d=2$ by adding the digits assigned to the edges 
$1,\dots,r-1$ in $\Z/d\Z$. 
Finite versions of such constructions
have applications in computer science (see e.g., M.~Luby~\&~A.~Wigderson~\cite{Luby}). 
\end{exmpl}

\section{Tail nesting} \label{sec:tailnest}

In this section we define and investigate tail nesting which is a 
specific way of nesting vertex copulas in order to shape tail dependencies.  

Let $\J$ be a grid decomposition of $\unitbox{r}$. The $\zero$-\emph{box} of $\J$
is the unique box in $\J$ which contains the origin~$\zero$.
For $\nu\in V$ we define the set $\J(\nu)$ of $\zero$-\emph{boxes with respect to} $\nu$ 
as the set of all $J\in\J$ such that 
$\pi^{F(\nu)}(J)$ is an $\zero$-box of $\pi^{F(\nu)}(\J)$ and $\pi^{F(\mu)}(J)$ 
is not an $\zero$-box of $\pi^{F(\mu)}(\J)$ 
for any $\mu\in F(\nu)\setminus\{\nu\}$. In this way we obtain a disjoint decomposition of $\J$, 
$$
\J=\bigcup_{\nu\in V} J(\nu)\,.
$$
Thus elements in $\J(\nu)$ project to the $\zero$-box in the decomposition of 
$F(\hat\nu)$ and $F(\hat\nu)$ is the face of maximal dimension having that property.

\begin{defi}[Tail nesting for grid copulas]
Suppose that $z\in C_k(u)$ is a vertex copula of dimension $r$ and that $c$ is a grid copula 
of order $k$ on $\unitbox{r}$ with grid~$\J$. Then $\tnest(z,c)$ is the grid copula 
obtained from $c$ by nesting
successively for each $\nu\in V$ the product measure  of $\pi^{F(\nu)}(z)$ with the uniform copula
$\unifc_{F(\nu)}$ as an element of  $C_k(u)$, that is
$$
\pi^{F(\nu)}(z)\times\unifc_{F(\nu)}\in C_k(u)\,,          
$$
into $\J(\nu)$. Thus $\tnest(z,c)$ is given by the consecutive nestings
\begin{equation} \label{eq:deftnest}
\nest\bigl(\pi^{F(\nu)}(z)\times\unifc_{F(\nu)},c,\J(\nu)\bigr)\,,\quad\nu\in V\,.
\end{equation}
The grid decomposition for $\tnest(z,c)$ is obtained from 
$\J$ by assigning to each box $J\in\J$ the vertex decomposition induced by the nesting. 
\end{defi}

\begin{figure}[ht] 
  \begin{center}
    \includegraphics[width=4.3cm]{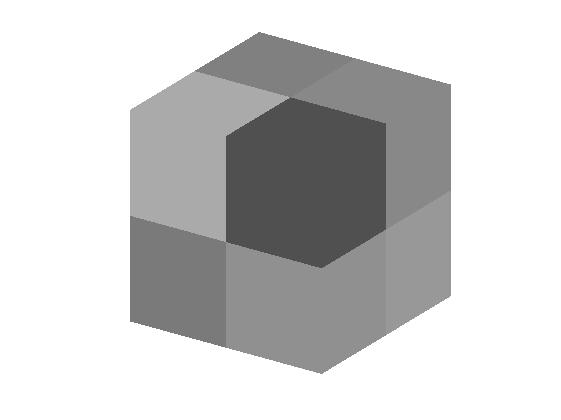}
    \includegraphics[width=4.3cm]{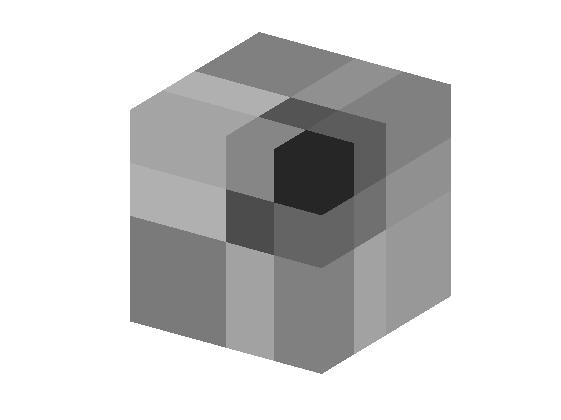}
    \includegraphics[width=4.3cm]{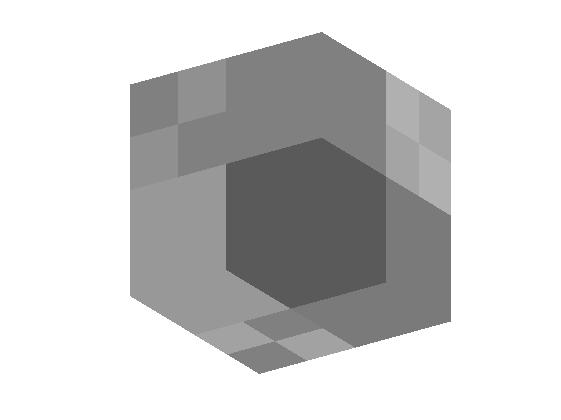}
    \caption{\label{fig:3d} Illustration of tail nesting a $3$-dimensional vertex copula $z$ 
    once into itself. The copula measure $z$ is pictured left. A front and back view of the 
    resulting box copula $\tnest(z,z)$ is illustrated middle and right, respectively. 
    The tail vertex $\zero$ is the upper front corner in the left and middle picture. 
    On the right hand side, the opposite vertex $\one$ corresponds to the lower front corner. 
    Grey levels are set according to the density of probability.}
 \end{center}
\end{figure} 

\begin{lem} \label{lem:pitnest}
In the setting of the above definition, $\tnest(z,c)$ is a grid copula of order $k$. 
Furthermore, tail-nesting commutes with projections along faces $F$ of $\unitbox{r}$,
$$
\pi^F\bigl(\tnest(z,c)\bigr)=\tnest\bigl(\pi^F(z),\pi^F(c)\bigr) 
$$
for the corresponding grid copulas. 
\end{lem}

\begin{proof}
It is a direct consequence of Lemma~\ref{lem:Ckinvundernest} that $\tnest(z,c)$ is a 
grid copula of order~$k$ provided $z,c$ are of order~$k$.

We prove now the second part of the statement.
For any face $F$ we denote here by $\unifc_F$ the uniform copula on $F$ with 
grid decomposition $\J_F\:=\pi^{F^c}(\J)$. 
Now let $F$ be a given face. Suppose that $\nu\in V\cap F^c$ and $I\in\J_{F^c}(\nu)$. 
Observe that $F(\nu)\subset F^c$. The elements in $\J$ which project along $F$ to $I$ 
are of the form $I\times K$, $K\in\J_F$ and in particular $K\subset F$. 
Observe that $I\times K\in\J(\mu)$ where $F(\mu)=F(\nu)\times F'$ with $F'\subset F$.  
Therefore we obtain
\begin{align*}
\pi^{F(\nu)}\Bigl(\pi^F(z)\Bigr)\times\unifc_{F(\nu)}
         &=\pi^F\Bigl(\pi^{F(\nu)}(z)\Bigr)\times\unifc_{F(\nu)} \\
         &=\pi^F\Bigl(\pi^{F(\mu)}(z)\times\unifc_{F(\mu)}\Bigr) \,.
\end{align*}
This implies that 
$
\tnest(\pi^F(z),\pi^F(c))|_I=\alpha\pi^F(\tnest(z,c)|_{I\times F})
$
for a scaling factor $\alpha\geq 0$. From the definition of nesting, $\alpha=1$.  
\end{proof}

In analogy to Example~\ref{exmpl:2d}, we can now investigate 
iterated tail nestings in dimension~$r$.
We pick a sequence $z_n$, $n=1,2\dots$ with $z_n\in C_k(u^\n)$, for $n\geq 1$. 
We define recursively grid copulas $c^n$ such that $c^1=z_1$ and 
$$
c^n=\tnest(z_n,c^{n-1})
$$
with grid decomposition $\J^n$. We denote  the limit copula $c^\infty$ also by $\tnest((z_n))$ and observe that
$c^\infty$ is a copula of order $k$. Lemma~\ref{lem:pitnest} implies that
\begin{equation} \label{eq:picinfty}
\pi^F\tnest\bigl((z_n)\bigr)=\tnest\bigl((\pi^F(z_n))\bigr)\quad\text{for any face\ } F.
\end{equation}
Note that the $\zero$-box of $c^n$ is $[0,d^\n]$ with 
\begin{equation} \label{eq:du}
d_i^\n=\prod_{l=1}^n u_i^{(l)}\quad i=1,\dots,r.
\end{equation}
The $\zero$-box of $\pi^{F(\nu)}(c^n)$ in $F(\hat\nu)$ is denoted by $[0,d^\n(\nu)]$
and given by
\begin{equation} \label{eq:dunu}
d_i^\n(\nu)=d_i^\n\text{\ if\ }\nu_i=0\quad\text{and}\quad d_i^\n(\nu)=0\text{\ if\ }\nu_i=1\,.
\end{equation}

\begin{rem}
Building a simulation algorithm for $c^n$ based on this definition and 
Algorithm~\ref{algo:nest} is not efficient. Observe that the nestings for 
$\nu\in V_{\leq k}$ 
do not change the measure and merely refine the decomposition. Including these 
trivial nestings formally in the definition has advantages in stating and 
proving properties of iterated tail nestings.
\end{rem}

\begin{algo}[for sampling tail nested copulas] \label{algo:tailnest}
Generate the samples based on Algorithm~\ref{algo:nest} and do the iterative nestings
only for $\zero$-boxes with respect to $\nu\in V_{>k}$. In the first iteration step one 
obtains a box copula which is illustrated in Fig.~\ref{fig:3d}. Then one needs to extend 
the definition of $\zero$-boxes to $\zero$-boxes with respect to $\nu\in V$ for the 
resulting box decomposition, which is easy. Proceeding in that way yields the required 
algorithm. In the first iteration, one ends up in the `otherwise'-routine of 
Algorithm~\ref{algo:nest} with a probability of $z_1(\cup_{\nu>k}\J(u^{(1)})(\nu))$ 
and conditioned on that, with a probability of $z_2(\cup_{\nu>k}\J(u^{(2)})(\nu))$ 
in the `nested otherwise'-routine and so on. Hence generating $N$ samples of the 
limit copula $c^\infty$ requires the sampling of approximately
$$
N':=N\Bigl(1+\sum_{n=1}^{\infty}\prod_{l=1}^n z_l\Bigl(\bigcup_{\nu>k}\J(u^{(l)})(\nu)\Bigr)\Bigr)
$$
random variates from a multinomial distribution for the 
vertices\footnote{In each `otherwise-routine', the number of vertices is $2^k$, $k\leq r$}.
If we have an upper bound $p_{\max}=\max_l(z_l(\cup_{\nu>k}\J(u^{(l)})(\nu)))<1$, then 
$$
N'\leq N(1-p_{\max})^{-1}
$$ 
and the algorithm converges. In practise, working with $p_{\max}$ sufficiently 
smaller than $1$ and with a finite sequence $(z_n)$ for shaping a desired 
tail behaviour should be sufficient.  
\end{algo}

We can coarsen the box decomposition $\J^n$ by building consecutively unions of two boxes, 
starting with boxes in $\J^n$, if they share a common face of codimension~$1$ and the 
probability density of $c^n$ on these boxes is the same, irrespective of the choice of
probabilities $z_l(\nu)$, $l=1,\dots,n$ and $\nu\in V$.  In this way we obtain the relevant 
box decomposition for Algorithm~\ref{algo:tailnest} and denote it by $\widetilde{\J}^n$. 
We observe that each box in 
\begin{equation} \label{eq:Jtilde}
\widetilde{\J}^n\setminus\bigcup_{\nu\in V_{>k}}\widetilde{\J}^n(\nu)
\end{equation}
is also contained in $\widetilde{\J}^m$ for $m\geq n$ and that $c^m$ restricted to 
these boxes is equal to $c^n$ restricted to these boxes. 

The theorem below summarises the main properties of iterated tail nestings
and demonstrates the flexibility one has when `shaping the tail' of $c^\infty$ 
by choosing appropriate $z_n$ with appropriate vertex decompositions given by $u^\n$. 
The constraint is given by Corollary~\ref{cor:Ck} and in particular 
Condition~\eqref{eq:boxcopcond} for the probability measure.

\begin{thm} \label{thm}
Let $V$ be the set of vertices of $\unitbox{r}$  and $S\colon\R^V\to\R^V$ denote the isomorphism given by~\eqref{eq:Sxmu}.
Assume that $(x_n)_{n\geq 1}$ is a sequence in $\R^V$ such that $S(x_n)\in C_k(u^\n)$ is a sequence of vertex copulas of order~$k$. Then $c^\infty:=\tnest\bigl((S(x_n))\bigr)$ is a copula of order~$k$ and satisfies
\begin{equation} \label{eq:cinftytail}
\bigl(\pi^{F(\nu)}c^\infty\bigr)([0,d^\n(\nu)])=\prod_{l=1}^n x_l(\nu),\quad \nu\in V,
\end{equation}
where $d_i^\n(\nu)=(1-\nu_i)\prod_{l=1}^n u_i^{(l)}$ for $i=1,\dots,r$. 
Furthermore, if $d^\n\to 0$ as $n\to\infty$ then $c^\infty$ is locally piecewise uniform
on $[0,1]^r\setminus \cup \F_{>k}$.
\end{thm}

\begin{proof}
By Proposition~\ref{prop:GkV}, in particular~(ii), and since $\tnest$ commutes with $\pi^F$, it is sufficient to show that~\eqref{eq:cinftytail} holds for $\nu=\zero$, i.e., that
$c^\infty([0,d^\n])=\prod_{l=1}^n x_l(\zero)$ where $d^\n=d^\n(\zero)$. This follows immediately from the definition of nesting.
  
Now assume that $d^\n\to 0$ as $n\to\infty$. As above, let $\J^n$ be the grid decomposition for $c^n=\tnest(z_n,c^{n-1})$ obtained in the process of iterated nestings. 
Then, given $\epsilon>0$, all elements of $\cup_{\nu>k}\J^n(\nu)$ are contained in the $\epsilon$-tube\footnote{The $\epsilon$-tube around a set $E$ is the set of all points with distance $<\epsilon$ to some point in $E$.} $W$ around $\cup \F_{>k}$ for all $n\geq n_0$, where $n_0$ is sufficiently large. Indeed, if the $\epsilon$-tube is taken with respect to the maximum norm, we may choose $n_0$ such that $\max_i\{d_i^\n\}<\epsilon$ for each $n\geq n_0$.
We obtain $c^\infty|_{\unitbox{r}\setminus W}=c^n|_{\unitbox{r}\setminus W}$ for $n\geq n_0$. Note that $c^n$, $n<\infty$, is piecewise uniform by~\eqref{eq:Jtilde}. Therefore $c^\infty$ is locally piecewise uniform in  $[0,1]^r\setminus\cup\F_{>k}$.      
\end{proof}

\begin{rem}We observe that \eqref{eq:cinftytail} also holds for the limit copula obtained by nesting iteratively $\tilde{c}_n=\nest(S(x_n),\tilde{c}_{n-1},\J^{n-1})$ with $\tilde{c}_1=S(x_1)$.
Tail nesting avoids those nestings which are not relevant for shaping the tail characteristic. Formally, this is done in \eqref{eq:deftnest} by averaging over 
those dimensions which do not matter for shaping the tail. 

In this context we notice the following. Suppose we start with a grid copula of order~$k$. We refine possibly the corresponding grid before
`tail-nesting' a sequence of vertex copulas of order~$k$ into the grid copula. In this way the original grid copula measure 
is merely modified in an arbitrarily small neighbourhood of $\cup\F_{>k}$, provided the refinement of the original grid was fine enough.  
\end{rem}

\section{Tail characteristics} \label{sec:tailchar}

Before we apply the theorem to construct copulas with certain tail characteristics, we investigate properties of any tail characteristic. 
To this end we assume that $c$ is any copula measure of order $k$, where $k\geq 1$, with $\tdeg_c<\infty$ and $\tcoef_c<\infty$. We denote $c_F:=\pi^{F^c}(c)$. For $s\in(0,1]$ we write
\begin{equation}  \label{eq:cFs}
c_F\bigr(s[0,1]^{\dim F}\bigl)=\tilde a(F,s)\,s^{\tdeg_c(F)}
\end{equation}
The condition for probability measures~\eqref{eq:boxcopcond} implies that
\begin{equation} \label{eq:tcharcond}
\sum_{F'\supset F}(-1)^{\dim F'-\dim F}\tilde a(F',s)\,s^{\tdeg_c(F')}\in [0,1]
\end{equation}
From this equation we obtain necessary condition for maps which are tail characteristics of copulas (see the remark below). 

As an application of the theorem we are going to state sufficient conditions. We have already introduced the properties \emph{non-decreasing} and \emph{increasing}
for maps $b\colon\F\to [0,\infty]$ in the introduction. 
We say $b\colon\F\to [0,\infty]$ is \emph{increasing at} $F\in\F$ and \emph{eventually constant at} 
$F$ if $b(F)<b(F')$ and $b(F)=b(F')$, respectively,  
for each $F'\in\F$ with $F\subset F'$ and $F\neq F'$.

If $b$ is increasing and $a$ is an accumulation point of the corresponding maps 
$\tilde a(\ ,s)$ as $s\to 0$,
we can see that the dominating summand in~\eqref{eq:tcharcond}
is $a(F,s)\,s^{\tdeg_c(F)}$, provided $a(F)>0$. 

\begin{rem}[Necessary conditions] \label{rem:tailchar}
From the above we obtain the following necessary conditions for 
maps 
$
a,b\colon\F\to (0,\infty)
$ 
such that $b=\tdeg_c$ for some copula $c$ of order $k$
and such that $a$ is an accumulation point of the corresponding maps 
$\tilde a(\ ,s)$ as $s\to 0$.
\begin{enumerate} 
\item $b$ is non-decreasing, i.e., $b(F')\geq b(F)$ for any $F'\supset F$,
\item $\bigl(a(F),b(F)\bigr)=(1,\dim F)$ for each $F\in\F^{\leq k}$. 
\item If $b$ is eventually constant at $F$, then 
      $
      \sum_{F'\supset F}(-1)^{\dim F'-\dim F}a(F')\geq 0
      $.
\end{enumerate}
If a pair $a,b$ satisfies (i)--(iii) for~$k$ we say that
$a,b$ satisfies $\NC_k$. 
\end{rem}

The tail characteristics of a copula $c$ restricted to the faces of dimension~$l$
provide a measure for \emph{tail dependencies of order~$l$}.
Given a map $\F\to\R$ as above we view it as well as a map
$V\to\R$ by composing it with the bijection $\nu\mapsto F(\hat\nu)$.

\begin{cor} \label{cor:tailapprox}
Let $a_m,b\colon\F\to (0,\infty)$, $m\geq 1$, satisfy~$\NC_k$ for some $k\geq 1$.  Suppose the sequence
$(a_m)_{m\geq 1}$ converges to some $a\colon\F\to (0,\infty)$. Assume in addition that $b$ is increasing.
Let $(s_m)_{m\geq 1}$ be a sequence in $(0,1)$ with $s_m\to 0$ as $m\to\infty$. Then, after passing to common subsequences, again denoted by 
$s_n,a_n$, $n\geq 1$, there exists a sequence $(z_n)_{n\geq 1}$ of vertex copulas of order~$k$, such that we have for $c^\infty:=\tnest((z_n)_{n\geq 1})$ that
\begin{equation} \label{eq:tailapprox}
c^\infty_F(s_n[0,1]^{\dim F})=a_n(F)\,s_n^{b(F)}\,.
\end{equation} 
\end{cor}

\begin{rem}
Let $c$ be any copula $c$ of order~$k$ with $\tdeg_c$ increasing. Choosing $s_m\to 0$, such that $a_m(F):=\tilde a(F,s_m)$ converges as $m\to\infty$ where $\tilde a(F,s)$ is as in~\eqref{eq:cFs}. 
Then the corollary states that after passing to a subsequence, $c^\infty_F(s_n[0,1]^{\dim F})=c_F(s_n[0,1]^{\dim F})$.
\end{rem}

\begin{proof}[Proof of Corollary~\ref{cor:tailapprox}]
For $a$ and $\delta>0$ we define open intervals
$
W_{\delta,a}:=(\min(a)-\delta,\max(a)+\delta)
$ 
and $W_{\delta,1}=(1-\delta,1+\delta)$.
We claim that there exists some $t_0\in(0,1)$ and $\delta\in (0,1)\cap (0,\min(a))$ such that for any $t<t_0$, any map
\begin{equation} \label{eq:h}
h\colon V\to W_{\delta,1}\cup W_{\delta,a}\,,
\quad\text{with}\quad 
h(\nu)=1\text{ for }\nu\in V_{\leq k}
\end{equation}
 the following is true: For 
$$x:=h\cdot t^b\in\R^V$$ 
with 
$
x(\nu)=h(\nu)t^{b(\nu)}
$
we have
$
z\:=S(x)\in C_k(t\one)\,.
$

The idea for the proof of this claim if already laid out in Remark~\ref{rem:tailchar}. 
Indeed, we need to check the condition for vertex copulas of order~$k$ in Corollary~\ref{cor:Ck}. Observe that $b$ and $h$ restricted to $V_{\leq k}$ are such that $z\in C_k(t\one)$ provided Condition~\eqref{eq:boxcopcond} for a probability measure is fulfilled. We estimate the respective expression 
\begin{equation} \label{eq:estimateSxnu}
S(x)(\nu)=\sum_{\mu\in F(\nu)}(-1)^{\mu-\nu}x(\mu) 		
\end{equation}
for $\nu\in V$ from below and from above by
\begin{align}
x(\nu)-\!\!\!\sum_{\mu\in F(\nu)\setminus\{\nu\}}\!\!x(\mu)   &\leq  S(x)(\nu)\leq
x(\nu)+\!\!\!\sum_{\mu\in F(\nu)\setminus\{\nu\}}\!\!x(\mu)\,,
\intertext{which is equivalent to}
h(\nu)t^{b(\nu)}-\!\!\!\sum_{\mu\in F(\nu)\setminus\{\nu\}}\!\!h(\mu)t^{b(\mu)} &\leq S(x)(\nu) \leq 
h(\nu)t^{b(\nu)}+\!\!\!\sum_{\mu\in F(\nu)\setminus\{\nu\}}\!\!h(\mu)t^{b(\mu)}\,. 
\label{eq:estimateSxnu3}
\end{align}
Since $h(\nu)>0$ and $b(\mu)>b(\nu)$ for each $\mu\in F(\nu)\setminus\{\nu\}$ by assumption on $b$ 
this proves that Condition~\eqref{eq:boxcopcond} holds for $\nu\neq\one$ if $t>0$ is 
small enough. Next observe that 
$$
S(x)(\one)=1-rt+\sum_{\nu\in V_{\geq 2}}(-1)^{\one-\nu}h(\nu)t^{b(\nu)}\in[0,1]
$$
provided $t$ is sufficiently small. 
Now we choose $t_0,\delta$ such that the above claim holds. 
Given the sequences $(s_m)$ and $(a_m)$ we choose the subsequences, again denoted by $(s_n)$ and $(a_n)$ such that 
\begin{align*}
	 t_1          &:=  s_1<t_0
&	 t_{n+1}      &:=  s_{n+1}/s_n< t_0\,, \\
	 h_1(\nu)     &:=  a_1(\nu)\in W_{\delta,a}
&  h_{n+1}(\nu) &:=  a_{n+1}(\nu)/a_n(\nu)\in W_{\delta,1} 
\end{align*}
for any $\nu\in V$. Then we set
$$
x_n:=h_n \cdot(t_n)^b\in C_k(t_n\one)
$$
and observe that 
$$
\prod_{l=1}^n t_l=s_n\quad\text{and}\quad
\prod_{l=1}^n x_n(\nu)=a_n(\nu)\prod_{l=1}^n (t_l)^{b(\nu)}
                      =a_n(\nu)(s_n)^{b(\nu)}\,.
$$
Now the corollary follows from the theorem.
\end{proof}


\begin{cor}
Suppose $a,b\colon\F\to (0,\infty)$ satisfy the condition $\NC_k$ for some $k\geq 1$. 
Assume further that 
$b$ is increasing.  Then there exists some $t\in [0,1]$ and a sequence $z_n\in C_k(t\one)$, such that 
$c^\infty=\tnest((z_n))$ satisfies
$
\tchar_{c^\infty}=(a,b)
$.
\end{cor}

\begin{proof}
Given $a,b$ as in the Corollary, we choose $t\in(0,1)$ and $\delta\in (0,1)$ sufficiently small such that 
\begin{equation}
(1+\delta')\cdot t^b  \in C_k(t\one) \quad\text{for each\ }\delta'\colon V\to(-\delta,\delta) 
\end{equation}  
We choose now a sequence $\delta_n\colon V\to(-\delta,\delta)$, $n\geq 1$ with $\delta_n(\nu)=0$ for $\nu\in V_{\leq k}$,     
such that 
$a_m:=\prod_{n=1}^m(1+\delta_n)\to a$ as $m\to\infty$. Then we set
$x_n:=(1+\delta_n)\cdot t^b$ for $n\geq 1$.   
We calculate next the tail degree of $c^\infty=\tnest((S(x_n)))$. Given $s\in (0,t)$ we determine 
$m\geq 2$ such that 
$
t^m\leq s < t^{m-1}\,.
$
We obtain for $\tau>0$ that
\begin{align}
     \frac{\bigl(\pi^{F(\nu)}c^\infty\bigr)(t^m    [0,\one(\nu)])}{t^{(m-1)\tau}} &\leq
     \frac{\bigl(\pi^{F(\nu)}c^\infty\bigr)(s      [0,\one(\nu)])}{s^{\tau}}     
   < \frac{\bigl(\pi^{F(\nu)}c^\infty\bigr)(t^{m-1}[0,\one(\nu)])}{t^{m\tau}}\,.  \notag
\intertext{Applying the theorem yields}
     t^{ \tau}\frac{a_m(\nu) t^{mb(\nu)}}{t^{m\tau}}                         &\leq
              \frac{\pi^{F(\nu)}\bigl(z^\infty\bigr)(s  [0,\one(\nu)])}{s^{\tau}}     
   < t^{-\tau}\frac{a_{m-1} a(\nu) t^{(m-1)b(\nu)}}{t^{(m-1)\tau}}         \notag
\end{align}
and therefore $\tdeg_{c^\infty}=b$. A continuity argument shows that there exists an adequate sequence $(\delta_n)$ such that $\tcoef_{c^\infty}=a$.  
\end{proof}

\begin{rem} \label{rem:limsupliminfbound}
We make the following observation in the previous proof. 
When nesting $z_n=S(x_n)$ with $x_n=(1+\delta_n)\cdot (t_n)^b$ where $\delta_n(\nu)\to 0$,
in order to construct $c^\infty=\tnest((z_n))$ with $\tdeg=b$,   
we can arrange that
$\liminf t_n\geq t_\infty>0$, $t_\infty$ depending on $b$. This enables us to obtain an upper bound for
$$
\limsup \tilde a(\,\cdot\,,s) / \liminf \tilde a(\,\cdot\,,s)
$$
depending only on $b$. Here $\tilde a$ is determined from $c^\infty$ as in~\eqref{eq:cFs}.

Furthermore, if we had $t_\infty=1$, then $\limsup \tilde a(\,\cdot\,,s)=\liminf \tilde a(\,\cdot\,,s)$ as $s\to 0$. In such a situation, the condition for the probability measure may become more difficult to control. Further below we will see that it can be easily controlled 
if the tail degree is equal to~$1$. 
\end{rem}

In the next application we weaken the condition that $b$ is increasing. 

\begin{cor} \label{cor:eventuallyconst}
Suppose that $a,b\colon\F\to (0,\infty)$ satisfy condition $\NC_k$ for some $k\geq 1$. 
Assume that $b$ is increasing or eventually constant at every $F\in\F$. 
Then, given any sequence in $(0,1]$ converging to~$0$,
there is a subsequence $(s_n)_{n\geq 1}$, a sequence $(z_n)_{n\geq 1}$ of vertex copulas of order~$k$ such that 
$$
\lim_{n\to\infty} \frac{c^\infty_F\bigl(s_n[0,1]^{\dim F}\bigr)}{(s_n)^{b(F)}}=a(F)\,.
$$
for $c^\infty=\tnest((z_n))$ and 
$\tdeg_{c^\infty}=b$.
\end{cor}

\begin{proof}
We proceed exactly as in the proof of Corollary~\ref{cor:tailapprox} with $(s_m)$ the originally given sequence and with $a_m=a$. To begin with we obtain 
the estimate~\eqref{eq:estimateSxnu3} for those~$\nu$ with~$b$ increasing at $F(\hat\nu)$ by arguing as above.

Setting $x_1=h_1\cdot(t_1)^b$ for $t_1<t_0$, we choose $h_1=a$ and it remains to show that $S(x_1)(\nu)\in[0,1]$ for those $\nu$ with~$b$ eventually constant at $F(\hat\nu)$. 
Observe that for these $\nu$, $b(\mu)=b(\nu)$ for each $\mu\in F(\nu)$. As $a$ satisfies~(iii) 
by assumption we see that $S(x_1)(\nu)\in[0,1]$ provided $t_0$ is sufficiently small. 
Without loss of generality assume $s_1=t_1$. 

Next we set $x_n=(t_n)^b$, $t_n=s_n/s_{n-1}$, $n>1$. 
We claim that $x_n\in C_k(t_n\one)$ after passing to an appropriate subsequence $(s_n)_{n\geq 1}$. Indeed, suppose $b$ is eventually constant at $F(\hat\nu)$ and $l=\dim F(\nu)>0$. Then we decompose the vertices of $F(\nu)$ along the lines of~\eqref{eq:Vr} into $(V\cap F(\nu))_l\cup\cdots\cup(V\cap F(\nu))_0$ and observe that 
the number of elements in $(V\cap F(\nu))_j$ is $\binom{l}{j}$. As 
$$
\sum_{j=0}^l \hbox{$\binom{l}{j}(-1)^j$}=0
$$
we see that $x_n(\nu)=0$. Hence $x_n$ satisfies the condition for a probability measure. 
For $z_n=S(x_n)$, $n\geq 1$, we see that $(z_n)$ has the desired properties.  
In view of Remark~\ref{rem:limsupliminfbound} we can achieve $\tdeg_{c^\infty}=b$. If the ratios
$s_n/s_{n+1}$ were not uniformly bounded from above we can enlarge the sequence $s_n$ by appropriate intermediate points in order to apply the arguments as in Remark~\ref{rem:limsupliminfbound}.
\end{proof}

Recall that the Clayton copula has tail dependence of degree~$1$ and
likewise the nested Clayton copulas which are described in~\cite{McN}. 
We conclude this section by investigating necessary and sufficient conditions 
for tail coefficients in case of tail degree~$1$.

\begin{rem}
Suppose $c$ is a copula of order~$k$ and $\tilde a(\,\cdot\, ,s)$ as in Remark~\ref{rem:tailchar}. 
Then any accumulation point $a$ of $\tilde a(\,\cdot\, ,s)$ satisfies condition~(ii) and~(iii) in that remark. 
If $c$ has tail dependence of degree~$k$, then (iii) is equivalent to
\begin{enumerate}
\item[(iii)'] 
$S(a)|_{F\cap V}\in\R^{F\cap V}$ is a probability measure for $F\in\F_k$ 
\end{enumerate}
with $a(\nu)=a(F(\hat\nu))$. Indeed, given $F\in\F^k$ choose $\nu\in V_k$ with $F(\hat\nu)=F$.
By assumption, $\tdeg_c(F')=k$ for any face $F'\supset F$ and 
thus $\tdeg_c$ is eventually constant at any $F'\supset F$.  
We recall that $F(\hat\mu)\supset F(\hat\nu)$ if and only if $\mu\in F(\nu)$. 
Given any $\nu'$ in $F\cap V$ we see from~(iii) that
$$
S(a)(\nu')=\sum_{\mu\in F(\nu')}(-1)^{\mu-\nu'}a(\mu)\geq 0\,.
$$
As $\sum_{\mu\in F(\nu)} S(a)(\mu)=a(\nu)=a(F(\hat\nu))=1$ this shows that~(iii)' 
holds. The other direction is now evident as well.  
\end{rem}

\begin{cor}
Let $a\colon\F\to [0,1]$ with $a(F)=1$ for $F\in\F^{\leq 1}$. Assume that $S(a)|_{F\cap V}$ is a probability measure for any $F\in\F_1$. 
Then there exists a sequence $(z_n)_{n\geq 1}$ of vertex copulas such that $c^\infty:=\tnest((z_n))$ has tail dependence of degree~$1$ and $\tcoef_{c^\infty}=a$. Moreover,  
$$
\lim_{s\to 0}\frac{c_F(s[0,1]^{\dim F})}{s}=a(F) 
$$ 
In other words, by means of tail nesting we can achieve any possible tail coefficients in case of
tail degree~$1$. As the above limit exists, the copulas $c_F$ with $\dim F=2$ have lower 
tail dependence $a(F)$. 
\end{cor}

\begin{proof}
Along the lines of the construction above, we set
$$
x_1:=a\cdot (t_1)^b
$$
with $b(\nu)=1$ for $\nu\neq\one$ and $b(\one)=0$. By the assumptions on $a$,
$S(x_1)\in C_1(t_1\one)$ provided $t_1>0$ is sufficiently small. Next we claim that
$
x:=t^b\in C_1(t\one)
$
for any $t\in(0,1)$. As $b$ is constant on $V_{>0}$ we need to verify only that
$S(t^b)(\one)\in [0,1]$. And indeed,
$$
S(t^b)(\one)=(t^0-t^1)+\sum_{\mu\in V}(-1)^{\mu-\one}t^1=(1-t)\in [0,1]
$$
for any $t\in(0,1)$.
Now choose $t_n\in(0,1)$, $n>2$, such that $t_n\to 1$ and set
$$
x_n:=(t_n)^b\,.
$$
Then $c^\infty:=\tnest\bigl((S(x_n))_{n\geq 1}\bigr)$ has the desired properties. 
\end{proof}

\begin{rem}
Other interesting examples of copulas with tail dependence of degree~$1$ are
$\tnest\bigl((S((1+\delta_n)(t_n)^b))_{n\geq 1}\bigr)$ for appropriate~$\delta_n$ with
and 
$$
\prod_{n=1}^\infty (1+\delta_n(\nu))=a(\nu)\,.
$$ 
In this way, we can 
control how the limits are approached, starting at an arbitrary $t_1\in (0,1)$.  
\end{rem}

\section{Change of coordinates} \label{sec:chcoord}

When studying tail characteristics for random variables $X_1,\dots,X_r$,
we could study for a given decreasing sequence $(0,1)\ni s_n\to 0$ and each $\nu\in V$ 
the asymptotic behaviour of   
\begin{equation} \label{eq:PX1Xr}
p_n(\nu)=P(X_i\leq Q_i(s_n)\text{ for }\nu_i=0) \quad\text{as }n\to\infty
\end{equation}
where $Q_i$ is the quantile function of $X_i$, i.e., the inverse of the cumulative distribution function $F_i\colon x\mapsto P(X_i\leq x)$.
We assume for simplicity that the $F_i$ are continuous and strictly increasing. 
As an application of Theorem~\ref{thm} we can use tail nesting in order to construct probability spaces with random variables $X_1,\dots,X_r$  where the asymptotic behaviour of all the functions in~\eqref{eq:PX1Xr} can be prescribed. 

The transformation to the uniform variables appearing in this context is just one of many possible transformation. One of the critical comments about the use of copulas is that 
``$\,(\dots)$ \emph{there is no particular mathematical or practical reason} $(\dots)$\,'' 
(Th. Mikosch~\cite{Mikosch}) for selecting this transformation.

More generally one may wish to look at the asymptotic behaviour of
\begin{equation} \label{eq:xi}
p(\xi^\n,\nu)=P(X_i\leq \xi_i^\n\text{ for }\nu_i=0) \quad\text{as }n\to\infty
\end{equation}
for a given sequence $\xi^\n\in\R^r$ such that $P(X_i\leq\xi_i^\n)>0$ and strictly decreasing to $0$.
As above, we can apply Theorem~\ref{thm} in order construct probability spaces together with random variables $X_1,\dots,X_r$ where
\begin{enumerate}
\item
the cumulative distribution function $F_i$ of $X_i$ is given, and 
\item the asymptotic behaviour of $p(\xi^\n,\nu)$ can be controlled simultaneously for all $\nu\in V$.
\end{enumerate} 
We can arrange the vertex copulas $x_n\in C_k(u^\n)$ in Theorem~\ref{thm} such that 
$$
F_i(\xi_i^\n)=d_i^\n\,.
$$ 
where $d_i^\n$ is as in~\eqref{eq:du}. The sequence $(u^\n)$ is determined by the sequence $(\xi^\n)$ and we can choose $x_n\in\R^V$ subject
to the conditions in Corollary~\ref{cor:Ck}. 
The transformation from $\xi^\n$ to $d^\n$ 
gives merely a nice coordinate system to carry out the geometric construction of nesting. 

\begin{exmpl} Consider a collection of Pareto-distributions   
$
F_i\colon (-\infty,-1]\to [0,1],\ s\mapsto (-s)^{-\alpha_i}
$ 
with $\alpha\in (0,\infty)^r$. Say we are choosing $\xi_i^\n=-(-t)^n$ for some $t>1$, i.e.,
we aim to control the asymptotic behaviour of the probabilities
$$
P(X_i\leq -(-t)^n\text{ for }\nu_i=0) \quad\text{as }n\to\infty\,.
$$
As $d_i^\n=(-t)^{-\alpha_in}$ we see that
$
u^\n=\bigl((-t)^{-\alpha_1},\dots,(-t)^{-\alpha_r}\bigr)=:u
$
which does not depend on $n$. 
We define now $x_n$ by
$$
x_n=(1+\delta_n)\cdot(-t)^{-b}
$$ 
where $b(\one)=0$, $\delta_n(\nu)=0$ for $\nu\in V_{\leq 1}$, $|\delta_n(\nu)|<\delta$, and 
$
\prod_n(1+\delta_n(\nu))=:a(\nu)\in (0,\infty)
$.
Furthermore, we require that
\begin{equation} \label{eq:montbmunu}
b(\nu) <b(\mu)\quad\text{for } F(\hat\nu)\subset F(\hat\mu)\text{ and }\nu\neq\mu\,,
\end{equation}
and
\begin{equation} \label{eq:minalphaib}
\max\{\alpha_i\,\mid\,\nu_i=0\}\leq b(\nu)\leq \sum_{\nu_i=0}\alpha_i,\quad\text{for }\nu\neq\one\,.
\end{equation}
We observe next that $S(x_n)=:z_n\in C_1(u)$, 
provided $|t|$ is sufficiently large. 
We impose the condition on the right hand side of~\eqref{eq:minalphaib} 
in order to ensure that the probabilities of $\zero$-boxes as approaching the origin $\zero$ 
are not smaller than in the case where the corresponding $X_i$ are independent. 
Note that 
for $\nu\in V_1$ the upper and lower bound in~\eqref{eq:minalphaib} 
are equal and their values consistent with the requirements in
Corollary~\ref{cor:Ck} for copulas (of order~$1$). For the probability measure $c^\infty:=\tnest((z_n))$ on $\unitbox{r}$ 
and $X_i(u)=F_i^{-1}(u_i)$, $u\in\unitbox{r}$
we obtain
$$
P(X_i\leq-(-t)^n\text{ for }\nu_i=0)\sim a(\nu) \bigl(-(-t)^{n}\bigr)^{-b(\nu)} \quad\text{as }n\to\infty\,.
$$
As in the previous section, we can weaken condition~\eqref{eq:montbmunu}
by dealing directly with constraints for $(\delta_n,b)$ imposed by Corollary~\ref{cor:Ck}. 
\end{exmpl}

\section*{Conclusion} 

The construction and examples described in this paper provide insights into 
a variety of asymptotic dependence structures of random variables. 
Tail nested copulas enable us to deal with tail dependencies of any order.
The behaviour of these copulas can be controlled along 
a sequence inside the unit $r$-cube which converges to the origin.

We believe that tail nested copulas are suitable for applications in risk management. 
They allow the risk modeller not only to take those dependencies into account 
which really matter in the specific application, but as well to generate 
corresponding stochastic samples numerically in an efficient manner.


\begin{thebibliography}{99}

\bibitem{CS} A.~Charpentier~\&~J.~Segers, \emph{Tails of Multivariate Archimedean Copulas}, {\tt arXiv:0901.1521 [math.PR]}, Jan.~2009.

\bibitem{Emb} P.~Embrechts, Copulas: A personal view, Department of Mathematics, ETH Z\"urich, 2007, to appear in \emph{Journal of Risk and Insurance}. 

\bibitem{FNL} G.~Fredricks et al., Copulas with fractal supports, \emph{Insurance: Mathematics and Economics}, \textbf{37} (2005), pp.~42--48.  

\bibitem{Luby} M.~Luby~\&~A.~Wigderson, \emph{Pairwise Independence and Derandomization}, 
Foundation and Trends in Theoretical Computer Science, \textbf{1} (4), pp 237--301, 2005.

\bibitem{McN} A.~McNeil, Sampling nested Archimedean copulas. \emph{Journal of Statistical Computation and Simulation}, \textbf{78}(6), 2008, pp.~567--581.

\bibitem{QRM} A.~McNeil, R.~Frey \& P.~Embrechts, \emph{Quantitative Methods in Risk Management}, Princeton Series in Finance, Princeton University Press, 2005.  

\bibitem{Mikosch}  Th.~Mikosch, Copulas: tales and facts, \emph{Extremes}, \textbf{9}, 3--20, 2006.

\bibitem{GMT} F.~Morgan, \emph{Geometric Measure Theory: A Beginner's Guide}, 
Academic Press, 1988. 

\bibitem{Nel} R.~Nelsen, \emph{An Introduction to Copulas}, Springer Verlag, 1999. 

\bibitem{Stoyanov} J.~Stoyanov, \emph{Counterexamples in Probability}, Second Edition, John Wiley \& Sons, 1987.

\bibitem{SP} D.~Stra\ss burger \& D.~Pfeifer, Dependence Matters!, Institut f\"ur Mathe\-ma\-tik, Carl von Ossietzky Universit\"at,  Oldenburg, presented at 36.~Internationales Astin-Kolloquium, ETH Z\"urich, Sep.~2005. 

\end{thebibliography}
\end{document}